# Reducing Network Traffic in Unstructured P2P Systems Using Top-k Queries[1]

Reza Akbarinia[1,2], Esther Pacitti[1], Patrick Valduriez[1]

[1]ATLAS group, INRIA and LINA, University of Nantes, France
[2]Shahid Bahonar University of Kerman, Iran
{FirstName.LastName@univ-nantes.fr, Patrick.Valduriez@inria.fr}

**Abstract.** A major problem of unstructured P2P systems is their heavy network traffic. This is caused mainly by high numbers of query answers, many of which are irrelevant for users. One solution to this problem is to use Top-k queries whereby the user can specify a limited number (*k*) of the most relevant answers. In this paper, we present FD, a (Fully Distributed) framework for executing Top-k queries in unstructured P2P systems, with the objective of reducing network traffic. FD consists of a family of algorithms that are simple but effective. FD is completely distributed, does not depend on the existence of certain peers, and addresses the volatility of peers during query execution. We validated FD through implementation over a 64-node cluster and simulation using the BRITE topology generator and SimJava. Our performance evaluation shows that FD can achieve major performance gains in terms of communication and response time.

## 1 Introduction

Peer-to-peer (P2P) systems adopt a completely decentralized approach to data sharing and thus can scale to very large amounts of data and users. Popular examples of P2P systems such as Gnutella [10] and KaaZa [12] have millions of users sharing petabytes of data over the Internet. Initial research on P2P systems has focused on improving the scalability of the unstructured systems, such as Gnutella and KaaZa, which rely on flooding. This work led to structured solutions that provide a distributed lookup mechanism to route search requests, *e.g.* CAN [15], CHORD [19], P-Grid [2] and FreeNet [7]. Although these designs can give better performance guarantees than unstructured systems, more research is needed to understand their trade-offs between autonomy, fault-tolerance, scalability, self-organization, etc. Meanwhile, the unstructured model which imposes no constraint on data placement and topology remains the most used today on the Internet.

A major problem of unstructured systems, which prevents them from being really scalable, is their heavy network traffic. Measurements in [17] have shown that although 95% of any two nodes are less than 7 hops away, the flooding based routing algorithm generates 330 TB/month in a Gnutella network with only 50,000 nodes. A main portion of this traffic is caused by the large amount of query answers, a lot of which may not be of interest to users. One obvious solution to this problem is to send the query only to the peers that are very close to the query originator [23], *e.g.* to those which are at most 3 hops a away. However, this significantly reduces the quality of results, in the sense that the user cannot get potentially "good" answers.

[1] Work partially funded by the ARA Massive Data of the Agence Nationale de la Recherche.

As another solution, we propose to use Top-k queries whereby the user can specify a number *k* and the system should return *k* of the most relevant answers. The degree of relevance (*score*) of the answers to the query is determined by a scoring function. Efficient techniques have been proposed for Top-k query processing in distributed systems [24] [25]. The algorithms typically use histograms, maintained at a central site, to estimate the score of databases with respect to the query and send the query to the databases that are more likely to involve top results. These techniques can somehow be used in super-peer P2P systems where super-peers maintain the histograms and perform query sending and result merging. However, because they rely on central information, these techniques no longer apply in unstructured systems.

In this paper, we present FD, a (Fully Distributed) framework for executing Top-k queries in unstructured P2P systems, with the goal of reducing network traffic. FD involves a family of algorithms that are simple but effective. FD has several salient features. First, it reduces significantly the communication cost of executing queries in unstructured systems. Second, its execution is completely distributed and does not depend on the existence of certain peers. Third, it addresses the volatility of peers during query execution and deals with situations where some peers leave the system before finishing query processing. We validated FD through a combination of implementation and simulation and the results show very good performance, in terms of communication and response time.

The rest of this paper is organized as follows. In Section 2, we make precise our assumptions and define the problem. In Section 3, we present the basic algorithm of FD, analyze its communication cost and propose techniques in order to reduce this cost. In Section 4, we address the volatility of peers by proposing algorithms extending the basic algorithm. Section 5 describes a performance evaluation of FD through implementation over a 64-node cluster and simulation using the BRITE topology generator [4] and SimJava [11]. Section 6 discusses related work. Section 7 concludes.

## 2 Problem Definition

In this section, we first give our assumptions regarding schema management and the P2P architecture. Then we can precisely state the problem we address in this paper.

In a P2P system, peers should be able to express queries over their own schema without relying on a centralized global schema as in data integration systems [20]. Several solutions have been proposed to support decentralized schema mapping. However, this issue is out of the scope of this paper and we assume it is provided using one of the existing techniques, *e.g.* [14], [20] and [3]. Furthermore, also for simplicity, we assume relational data.

We assume that the P2P system is unstructured, so the only requirement is that each peer knows some other peers, its neighbors, to communicate. In an unstructured P2P environment, there are two major aspects that make query processing difficult:

- **No centralized information**: there is no node for keeping global information about the data shared by all peers. Each peer keeps its own shared data and has no idea about the data shared by the other peers. The only thing that a peer knows is its neighbors' addresses.
- **Dynamicity:** peers are very dynamic and can join or leave the system at any time. During the execution of the query, some participating peers may thus leave the system.

Now we can define the problem as follows. Let *Q* be a Top-k query, *i.e.* the user is interested to receive *k* top answers to *Q*. Let TTL (Time-To-Live) determine the maximum hop distance which the user wants her query be sent. Let *D* be the set of all data items (*i.e.* tuples) that can be accessed through *ttl* hops in the P2P system during the execution of *Q*. Let *Sc(d, Q)* be a scoring function that denotes the score of relevance of a data item $d \in D$ to *Q*. Our goal is to find the set $T \subseteq D$, such that:

$|T| = k$ and $\forall d_1 \in T,\ \forall d_2 \in (D - T)$ then $Sc(d_1, Q) \geq Sc(d_2, Q)$

while minimizing the communication cost.

# 3 Top-k Query Processing

In this section, we first present the basic algorithm of FD. Then, we analyze its communication cost and propose some techniques for reducing it. For simplicity, we assume no dynamicity of peers which means that all peers remain reachable during query processing. This assumption will be relaxed in the next section.

## 3.1 Basic Algorithm

The algorithm starts at the *query originator*, the peer at which a user issues a Top-k query *Q*. The query originator performs some initialization. First, it sets *TTL* with a value which is either specified by the user or default. Second, it gives *Q* a unique identifier, denoted by *QID*, which is made of a unique peer-ID and a query counter managed by the query originator. Peers use *QID* to distinguish between new queries and those received before. After initialization, the query originator triggers the sequence of the following four phases (see the example in Figure 1): query forward, local query execution, merge-and-backward, and data retrieval.

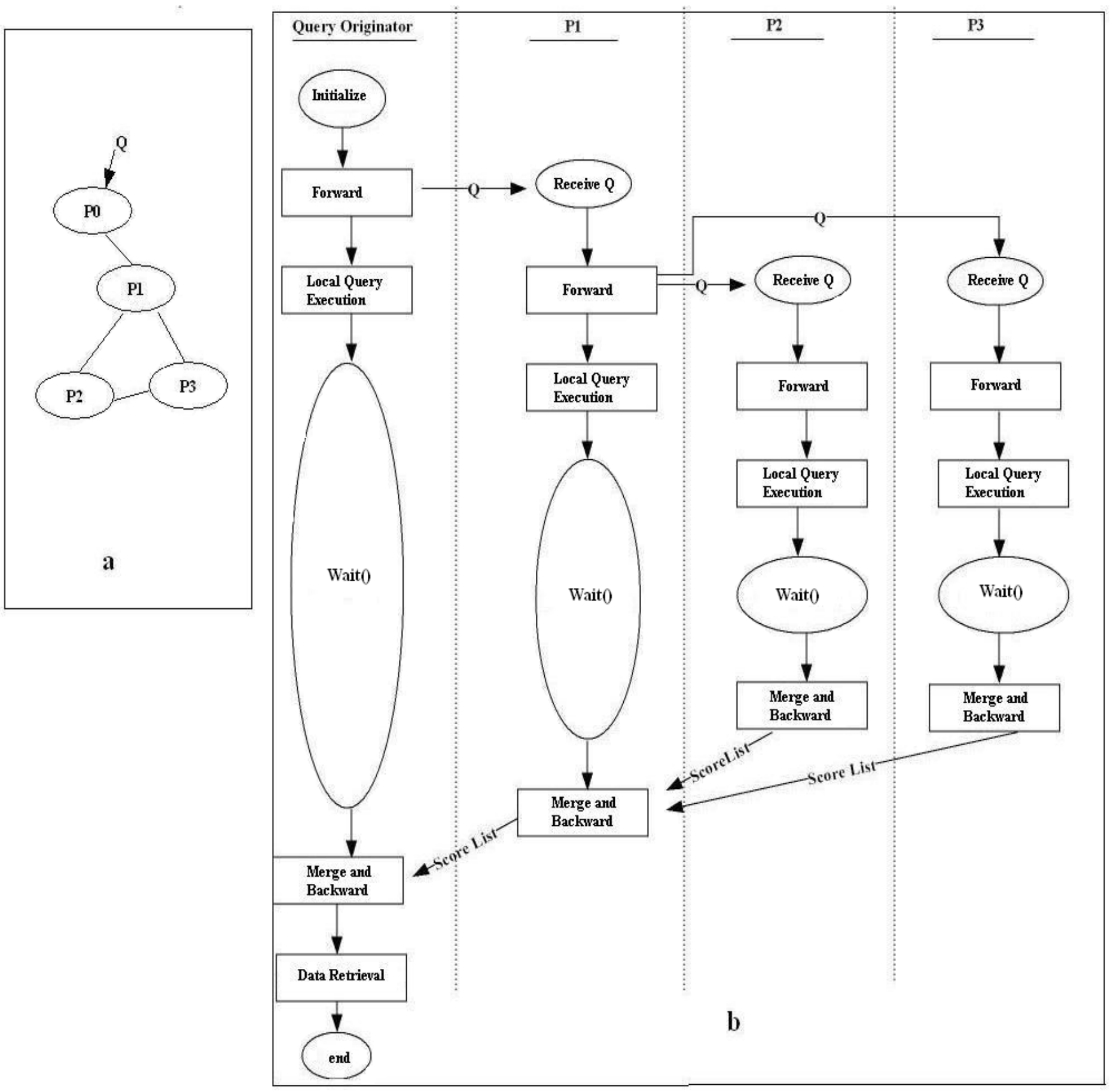

**Fig. 1.** A sample P2P system (a) and a sequence diagram of the basic algorithm of FD (b)

**Query Forward**

*Q* is included in a message that is broadcast by the query originator to its reachable neighbors. Each peer *p* that receives the message including *Q* performs the following steps.

1. Check *QID*: if *Q* has been already received, then discard the message else save the address of the sender as the *parent* of *p*.
2. Decrement *TTL* by one: if *TTL > 0*, make a new message including *Q*, *QID*, new *TTL* and the query originator's address and send the message to all neighbors (except parent).

**Local Query Execution**

After sending *Q* to its neighbors, *p* executes *Q* locally, *i.e.* accesses the local data items that match the query predicate, scores them using a scoring function, selects the *k* top data items and saves them as well as their scores locally. For scoring the data items, we can use one of the scoring functions proposed for relational data, *e.g.* Euclidean function [6][5]. These functions require no global information and can score peer's data items only using local information. The scoring function can also be specified explicitly by the user.

After selecting the *k* local top data items, *p* must wait to receive its neighbors' score-lists before starting the next phase. However, since some of the neighbors may leave the P2P system and never send a score-list to *p*, we must set a limit for the wait time. We compute *p*'s wait time using a cost function based on TTL, network dependent parameters and *p*'s local processing parameters. We provide the details of this cost function in Appendix A. If the cost function is inaccurate (for some peers), then it may happen that score-lists get received after the expiration of the wait time. We will deal with this situation in Section 4.

**Merge-and-Backward**

After its wait time has expired, *p* merges its *k* local top scores with those received from its neighbors and sends the result to its *parent* (the peer from which it received *Q*) in the form of a *score-list*. In order to minimize network traffic, we do not "bubble up" the top data items (which could be large), only their scores and addresses. A score-list is simply a list of *k* couples *(a, s)*, such that *a* is the address of the peer owning the data item and *s* its score. Thus, *p* performs the following steps:

1. Merge the score-lists received from the neighbors with its local *k* top scores and extracting the *k* top scores (along with the peers' addresses).
2. Send the *merged score-list*, which contains the *k* highest scores (and peers' addresses) extracted from local top scores and those received from the neighbors, to its parent.

**Data Retrieval**

By the three first phases, the merged score-lists containing top scores are bubbled up to the query originator. After the query originator has produced its merged score-list, which is called the *final score-list*, and which is gained by merging its *k* local top scores with the merged score-lists received from its neighbors, it directly retrieves the *k* top data items from the peers in the list as follows. For each peer address *a* in the final score-list:

1. Determine the number of times *a* appears in the final score-list, say *m* times.
2. Ask the peer at *a* to return its *m* top scored items.

Formally, consider the final score-list $L_f$ which is a set of at most *k* couples *(a, s)*, in this phase for each $a \in Domain(L_f)$, the query originator determines $T_a = \{s \mid (a, s) \in L_f\}$ and asks the peer at *a* to return $|T_a|$ of its top scored items.

## 3.2 Analysis of Communication Cost

In this section, we analyze our basic algorithm's communication cost. As we will see, it is not very high. We also propose strategies to reduce it. We measure the communication cost in terms of number of messages and number of bytes which should be transferred over the network in order to execute a query by our algorithm. The messages transferred can be classified as: 1) *forward messages*, for forwarding the query to peers. 2) *backward messages*, for returning the score-lists from peers to the query originator. 3) *retrieve messages*, to request and retrieve the $k$ top results. We first present a model representing the peers that collaborate on executing our algorithm, and then analyze the communication cost of backward, retrieve and forward messages.

**Model**

Let $P$ be the set of the peers in the P2P system. Given a query $Q$, let $P_Q \subseteq P$ be a set containing the query originator and all peers that receive $Q$. We model the peers in $P_Q$ and the links between them by a graph $G(P_Q, E)$ where $P_Q$ is the set of *vertices* in $G$ and $E$ is the set of the *edges*. There is an edge $p$-$q$ in $E$ if and only if there is a link between the peers $p$ and $q$ in the P2P system. Two peers are called *neighbor*, if and only if there is an edge between them in $G$. The number of neighbors of each peer $p \in P_Q$ is called the *degree of p* and is denoted by $d(p)$. The average degree of peers in $G$ is called the *average degree of G* and is denoted by $d(G)$. The average degree of $G$ can be computed as $d(G) = (\sum_{p \in P_Q} d(p)) / |P_Q|$

During the execution of our algorithm, $p \in P_Q$ may receive $Q$ from some of its neighbors. The first peer, say $q$, from which $p$ receives $Q$, is the *parent* of $p$ in $G$, and thereby $p$ is a *child* of $q$. A peer may have some neighbors that are neither its parent nor its children.

**Backward Messages**

In the Merge-and-Backward phase, each peer in $P_Q$, except the query originator, sends its merged score-list to its parent. Therefore, the number of backward messages, denoted by $m_{bw}$, is $m_{bw} = |P_Q| - 1$.

Let $L$ be the size of each element of a score-list in bytes (*i.e.* the size of a score and an address), then the size of the score-list is $k \times L$, where $k$ is the number of top results specified in $Q$. Since the number of score-lists transferred by backward messages is $|P_Q| - 1$, then the total size of data transferred by backward messages, denoted by $b_{bw}$, can be computed as $b_{bw} = k \times L \times (|P_Q| - 1)$. If we set $L=10$, *i.e.* 4 bytes for the score and 6 bytes for the address (4 bytes for IP address and 2 bytes for the port number), then $b_{bw} = k \times 10 \times (|P_Q| - 1)$.

Let us show with an example that $b_{bw}$ is not significant. Consider that $10{,}000$ peers receive $Q$ (including the query originator), thus $|P_Q| = 10{,}000$. Since users are interested in a few results and $k$ is usually small, we set $k=20$. As a result, $b_{bw}$ is less than 2 megabytes. Compared with the tens of megabytes of music and video files, which are typically downloaded in P2P systems, this is small.

**Retrieve Messages**

By retrieve messages, we mean the messages sent by the query originator to request the $k$ top results and the messages sent by the peers owning the top results to return these results. In the Data Retrieval phase, the query originator sends at most $k$ messages to the peers owning the top results (there may be peers owning more than one top result) for requesting their top results and these peers return their top results by at most $k$ messages. Therefore, the number of retrieve messages, denoted by $m_{rt}$, is $m_{rt} \leq 2\times k$.

**Forward Messages**

Forward messages are the messages that we use to forward $Q$ to the peers. According to the basic design of our algorithm, each peer in $P_Q$ sends $Q$ to all its neighbors except its parent. Let $p_o$ denote the query originator. Consider the graph $G(P_Q, E)$ described before, each $p\in( P_Q - \{p_o\})$, sends $Q$ to $d(p)-1$ peers, where $d(p)$ is the degree of $p$ in $G$. The query originator sends $Q$ to all of its neighbors, in other words to $d(p_o)$ peers. Then, the sum of all forward messages $m_{fw}$ can be computed as

$$m_{fw} = (\sum_{p\in (P_Q - \{p_o\})} (d(p) - 1)) + d(p_o)$$

We can write $m_{fw}$ as follows :

$$m_{fw} = (\sum_{p\in P_Q} (d(p) - 1)) + 1 = (\sum_{p\in P_Q} (d(p)) - |P_Q| + 1$$

Based on the definition of $d(G)$, $m_{fw}$ can be written as $m_{fw} = (d(G) -1)\times|P_Q|+1$, where $d(G)$ is the average degree of $G$. According to the measurements in [16], the average degree of Gnutella is 4. If we take this value as the average degree of the P2P system, *i.e.* $d(G)=4$, we have $m_{fw} = 3\times|P_Q|+1$. From the above discussion, we can derive the following lemma.

**Lemma 1**: The number of forward messages in the basic algorithm is $(d(G) - 1)\times|P_Q|+1$.

**Proof:** Implied by the above discussion. □

To determine the minimum number of messages necessary for forwarding $Q$, we prove the following lemma.

**Lemma 2**: The lower bound of the number of forward messages for sending $Q$ to all peers in $P_Q$ is $|P_Q| - 1$.

**Proof:** For sending $Q$ to each peer $p\in P_Q$, we need at least one forward message. Only one peer in $P_Q$ has $Q$, *i.e.* the query originator, thus $Q$ should be sent to $|P_Q| - 1$ peers. Therefore, we need at least $|P_Q| - 1$ forward messages to send $Q$ to all peers in $P_Q$.□

Thus, the number of forward messages in the basic algorithm is far from the lower bound.

## 3.3 Reducing the Number of Messages

We can still reduce the number of forward messages using the following strategies. 1) sending $Q$ across each edge only once. 2) Sending with $Q$ a list of peers that have received it. 3) using statistics to send $Q$ to only a subset of neighbors, which are more expected to return top results.

**Sending *Q* across each edge only once**

In graph $G$, there may be many cases that two peers $p$ and $q$ are neighbors and none of them is the parent of the other, *e.g.* two neighbors which are children of the same parent. In these cases, in the basic form of our algorithm, both peers send $Q$ to the other, *i.e.* $Q$ is sent across the edge $p$-$q$ twice. We develop the following strategy to send $Q$ across an edge only once.

**Strategy 1**: When a peer $p$ receives $Q$, say at time $t$, from its parent (which is the first time that $p$ receives $Q$ from), it waits for a random, small time, say $\lambda$, and then sends $Q$ only to the neighbors which $p$ has not received $Q$ from them before $t + \lambda$.

**Lemma 3**: With a high probability, the number of forward messages with Strategy 1 is reduced to $d(G)\times|P_Q| / 2$.

**Proof:** Since $\lambda$ is a random number and different peers generate independent random values for $\lambda$, the probability that two neighbors send $Q$ to each other simultaneously is very low. Ignoring the cases where two neighbors send $Q$ to the other simultaneously, with Strategy 1, $Q$ is sent across an edge only once. Therefore, the number of forward messages can be computed as $m_{fw} = |E|$. Since $|E| = d(G)\times|P_Q|/2$, then $m_{fw} = d(G)\times|P_Q|/2$. □

Considering $d(G)=4$ (similar to [16]), the number of forward messages is $m_{fw} = 2\times|P_Q|$.

With Strategy 1, $m_{fw}$ is closer to the lower bound than the basic form of our algorithm. However, we are still far from the lower bound. By combining Strategy 1 and another strategy, we can reduce the number of forward messages much more.

### Attaching to each forward message a list of peers that have received *Q*

Even with Strategy 1, between two neighbors, which are children of the same parent $p$, one forward message is sent although it is useless (because both of them have received $Q$ from $p$). If $p$ attaches a list of its neighbors to $Q$, then its children can avoid sending $Q$ to each other. Thus, we propose a second strategy.

**Strategy 2**: Before sending $Q$ to its neighbors, a peer $p$ attaches to $Q$ a list containing its Id and the Id of its neighbors and sends this list along with $Q$. Each peer that receives the $Q$'s message, verifies the list and does not send $Q$ to the peers involved in the list.

**Theorem 1**: By combining Strategy 1 and Strategy 2, with a high probability, the number of forward messages is less than $d(G)\times|P_Q|/2$.

**Proof:** With Strategy 2, two neighbors, which have the same parent, do not send any forward message to each other. If we use Strategy 1, with a high probability at most one forward message is sent across each edge. Using Strategy 2, there may be some edges such that no forward message is sent across them, *e.g.* edges between two neighbors with the same parent. Therefore, by combining Strategy 1 and Strategy 2, the number of forward messages is $m_{fw} \le |E|$, and thus $m_{fw} \le d(G)\times|P_Q|/2$. □

Considering $d(G)=4$, the number of forward messages is $m_{fw} \le 2\times|P_Q|$.

### Using statistics to reduce the messages

If the peers send $Q$ to only a subset of their neighbors, which are more likely to return the $k$ top answers, then we can significantly reduce the number of messages, including the forward and backward messages. However, the peers need some statistics to select the best neighbors. For this, each peer $p$ keeps some statistics about the content of the score-lists returned by every neighbor. These statistics can include: 1) The number of scores returned by the neighbor that are in the $p$'s merged score-list. 2) The position of the greatest score, returned by the neighbor, in the $p$'s merged score-list. Using these statistics, we can develop a number of heuristics to send *Q* to the neighbors, which are more likely to return the top results. These heuristics include:

- Do not send *Q* to the neighbors which none of their scores was in the merged score-list in the previous execution of *Q*.
- Send *Q* to the neighbors for which at least *x* percent (*e.g.* $x=10$) of their scores was in the merged score-list.
- Send *Q* to the neighbors which the position of their highest returned score in the merged score-list was lower than $z \times n$, where $z \leq 1$, *e.g.* $z=0.80$. With this heuristic, we select the neighbors that have previously returned better top scores.

Using the above heuristics, we can reduce the number of messages, but this may decrease the quality of the results. However, our experiments (see Section 5) show that, by selecting a good heuristic and well adjusting the parameters, we can achieve a significant reduction in the messages without a significant loss of quality.

## 4 Dealing with Peers' Dynamicity

In Section 3, we proposed FD's basic algorithm assuming no dynamicity of peers. Obviously, P2P systems are very dynamic, and it may well happen that some peers leave the system (or fail) at any time during query processing. Furthermore, peers can suddenly take more time than expected to respond. This can create the following problems to our basic algorithm: late reception of score-lists by a peer, after its wait time has expired; peers becoming inaccessible in the Merge-and-Backward phase; and peers that hold top data items becoming inaccessible in the Data Retrieval phase. In this section, we deal with these problems and propose extensions to the basic algorithm.

### 4.1 Late Reception of Score-Lists

In FD's basic algorithm, each peer *p*, after its wait time has expired, merges its top local scores with the score-lists received from its neighbors and sends the result as a score-list to its parent. However, *p* may underestimate its wait time which is based, among other parameters, on local processing parameters of other peers. Thus, it may happen that score-lists of some neighbors arrive late, *i.e.* after *p* has sent its score-list to its parent. We could simply ignore these late score-lists and discard them. However, they may refer to answers which are highly relevant to the user's query. Thus, we propose that *p* sends the late-score lists as an *urgent score-list* to its parent. The urgent score-lists should be bubbled up without wait until arriving to a peer of which wait time has not expired. When a peer *q* receives an urgent score-list, it performs the following actions:

- If *q* is not the query originator: if it has already sent its merged score-list to its parent, sends the urgent score-list immediately to its parent too; else, deals with the urgent score-list as any other received score-list.

- If *q* is the query originator: if it is in the Data Retrieval phase, discards the urgent score-list; else, deals with the urgent score-list as any other received score-list.

Therefore, using urgent score-lists, we can save late score-lists from being discarded (except those that reach to the query originator in the Data Retrieval phase), and this increases the accuracy of the algorithm's answer to the Top-k query.

### 4.2 Peers Inaccessible in the Merge-and-Backward Phase

In the Merge-and-Backward phase, each peer *p* sends a score-list to its parent. And it may happen that *p*'s parent is inaccessible (*e.g.* it has left the system). We could simply ignore *p*'s score-list and have *p* discard it. But we can find alternative paths to backward *p*'s core list. We propose the following strategy:

- If *p* has a neighbor, say *q*, which is not *p*'s child (*i.e. p* is not the first peer from which *q* has received *Q*), send to *q* the score-list as an urgent score-list. The urgent score-list will be bubbled up rapidly until it reaches a peer of which wait time has not expired.
- If *p* has not such a neighbor, send the score-list directly to the query originator which can deal with this score-list like any other score-list received from its neighbors. Recall from Section 3 that the address of the query originator is communicated to all peers along with *Q*.

Using this strategy, if the parent of a peer *p* leaves the system during query execution, the score-list of *p* is never lost.

### 4.3 Top Data Items Inaccessible in the Data Retrieval Phase

In the Data Retrieval phase, the peers which hold the top data items in the final-score list need be accessed by the query originator. However, it may happen that one or more of those peers are inaccessible, *e.g.* because of leaving the system, thus hampering the production of the complete final result. One simple way to deal with this is to produce an incomplete result, with less than k top items. A better solution is to increase k before starting the first phase to compensate the inaccessible data items. But, how many should we add to *k*? We can answer this question if we know a little more about peers' accessibility. Let $P<1$ be the probability that any top data item belonging to the final score-list be inaccessible, the following lemma gives us an answer:

**Lemma 4:** If we increase *k* to $k / (1 - P)$, then the expected number of accessible top data items in the Data Retrieval phase is *k*.

**Proof**: If the requested list size is *x*, then $x(1-P)$ items are expected to be accessible. Solving $x(1-P)=k$, the required value of *x* is $x=k/(1- P)$. □

Therefore, to compensate the inaccessible top data items in the Data Retrieval phase, before sending *Q* to its neighbors, the query originator can set *k* as $k / (1 - P)$. The query originator can use the statistics gathered from previous query executions for estimating *P*.

## 5 Performance Evaluation

We evaluated the performance of FD through implementation and simulation. The implementation over a 64-node cluster was useful to validate our algorithm and calibrate our simulator. The simulation allows us to study scale up to high numbers of peers (up to 10,000 peers).

The rest of this section is organized as follows. In Section 5.1, we describe our experimental and simulation setup, and the algorithms used for comparison. In Section 5.2, we evaluate the response time of FD. We first present experimental results using the implementation of FD and two other baseline algorithms on a 64-node cluster, and then we present simulation results on the response time using various parameters: number of peers, effect of latency, and bandwidth. Section 5.3 presents the evaluation of communication cost, and Section 5.4 evaluates the accuracy of FD w.r.t. the dynamicity of P2P systems.

### 5.1 Experimental and Simulation Setup

For our implementation and simulation, we used the Java programming language, the SimJava package and the BRITE universal topology generator.

SimJava [11] is a process based discrete event simulation package for Java. Based on a discrete event simulation kernel, SimJava includes facilities for representing simulation objects as animated icons on screen. A SimJava simulation is a collection of entities each running in its own thread. These entities are connected together by ports and can communicate with each other by sending and receiving event objects.

BRITE [4] has recently emerged as one of the most promising universal topology generators. The objective of BRITE is to produce a general and powerful topology generation framework. Using BRITE, we generated topologies similar to those of P2P systems and we used them for determining the linkage between peers in our tests.

We first implemented FD in Java on the largest set of machines that was directly available to us. The cluster has 64 nodes connected by a 1-Gbps network. Each node has an Intel Pentium 2.4 GHz processor, and runs the Linux operating system. We make each node act as a peer in the P2P system. To have a P2P topology close to real P2P overlay topologies, we determined the peer neighbors using the topologies generated by the BRITE universal topology generator [4]. Thus, each node only is allowed to communicate with the nodes that are its neighbors in the topology generated by BRITE.

To study the scalability of FD far beyond 64 peers and to play with various performance parameters, we implemented a simulator using SimJava. To simulate a peer, we use a SimJava entity that performs all tasks that must be done by a peer for executing FD. We assign a delay to communication ports to simulate the delay for sending a message between two peers in a real P2P system. For determining the links between peers, we used the topologies generated by BRITE.

In all our tests, we use the following simple query as workload:

SELECT R.data FROM R ORDER BY R.score
STOP AFTER *k*

Each peer has a table *R(score, data)* in which attribute *score* is a random real number in the interval *[0..1]* with uniform distribution, and attribute *data* is a random variable with normal distribution with a mean of 1 (kilo bytes) and a variance of 64. Attribute *score* represents the score of data items and attribute *data* represents (the description of) the data item that will be returned back to the user as the result of query processing. The number of tuples of *R* at each peer is a random number (uniformly distributed over all peers) greater than 1000 and less than 20,000.

The simulation parameters are shown in Table 1. Unless otherwise specified, the latency between any two peers is a normally distributed random number with a mean of 200 (ms) and a variance of 100. The bandwidth between peers is also a random number with normal distribution with a mean of 56 (kbps) and a variance of 32. Since users are usually interested in a small number of top results, we set *k=20*.

The simulator allows us to perform tests up to *10,000* peers, after which the simulation data no longer fit in RAM and makes our tests difficult. This is quite sufficient for our tests. Therefore, the number of peers of P2P system is set to be *10,000*, unless otherwise specified. In all tests, *TTL* is set as the maximum hop-distance to other peers from the query originator, thus all peers of the P2P system can receive *Q*. We observed that in the topologies with *10,000* nodes, with *TTL=12* all peers could receive *Q*. Our observations correspond to those based on experiments with the Gnutella network [16]; for instance, with *50,000* nodes, the maximum hop-distance between any two nodes is *14*.

**Table 1.** Simulation parameters

| Parameter | Values |
|---|---|
| Bandwidth | Normally distributed random, Mean = 56 Kbps, Variance = 32 |
| Latency | Normally distributed random, Mean = 200 ms, Variance = 100 |
| Number of peers | 10,000 peers |
| TTL | Large enough such that all of peers can receive the query |
| *K* | 20 |
| Result data items size | Normally distributed random, Mean = 1 KB, Variance = 64 |

In our simulation, we compare FD with two other algorithms. The first algorithm is a *centralized* algorithm, which we denote as *CN*, which works as follows. Each peer receiving *Q* sends its *k* top relevant data items directly to the query originator. The query originator merges the received results and extracts the *k* overall top scored data items from them. The second algorithm is an optimized version of CN, which we denote as *CN**, by which the peers return directly to the query originator only their score-lists (not data items).

### 5.2 Response Time

#### Scale up

In this section, we investigate the scalability of FD. We use both our implementation and our simulator to study response time while varying the number of peers. The re-sponse time includes local processing time and data transfers, *i.e.* sending query messages, score-lists and data items.

Using our implementation over the cluster, we ran experiments to study how response time increases with the addition of peers. Figure 2 shows excellent scale up of FD since response time logarithmically increases with the addition of peers until 64. Using simulation, Figure 3 shows the response times of the three algorithms with a number of peers increasing up to 10000 and the other simulation parameters set as in Table 1.

FD always outperforms the two other algorithms and the performance difference increases significantly in favor of FD as the number of peers increases. The main reason for FD's excellent scalability is its fully distributed execution. With CN and CN*, a central node, *i.e.* the query originator, is responsible for query execution, and this creates two problems. First, the central node becomes a communication bottleneck since it must receive a large amount of data from other peers that all compete for bandwidth. Second, the central node becomes a processing bottleneck, as it must merge many answers to extract the *k* top results.

Overall, the experimental results correspond with the simulation results. However, the response time gained from our experiments over the cluster is a little better than that of simulation because the cluster has a high-speed network.

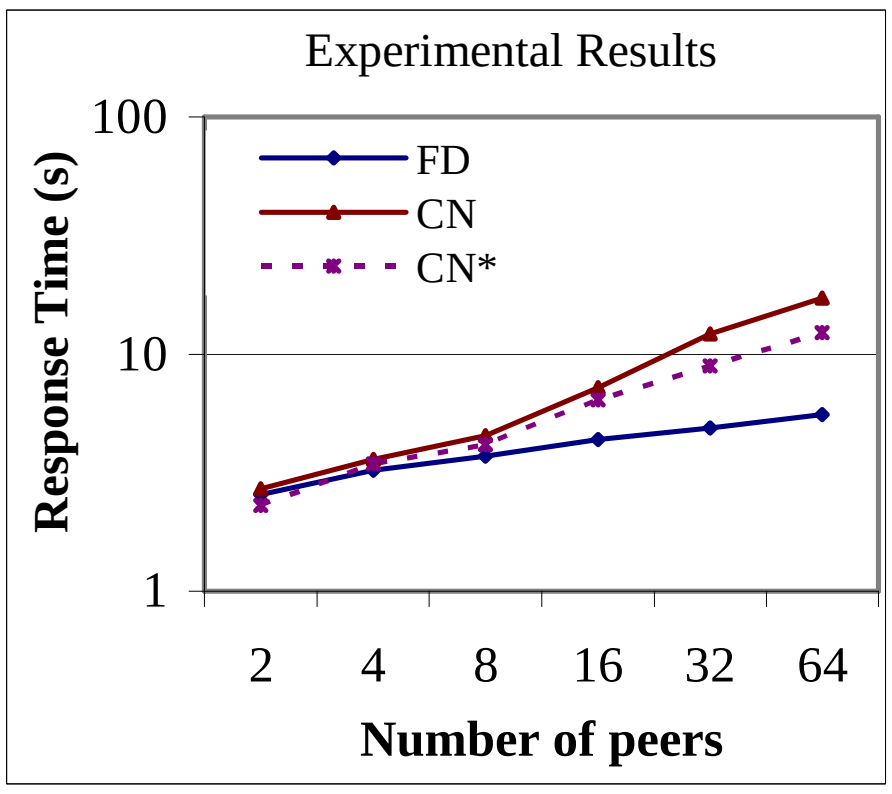

**Fig. 2.** Response time vs. number of peers

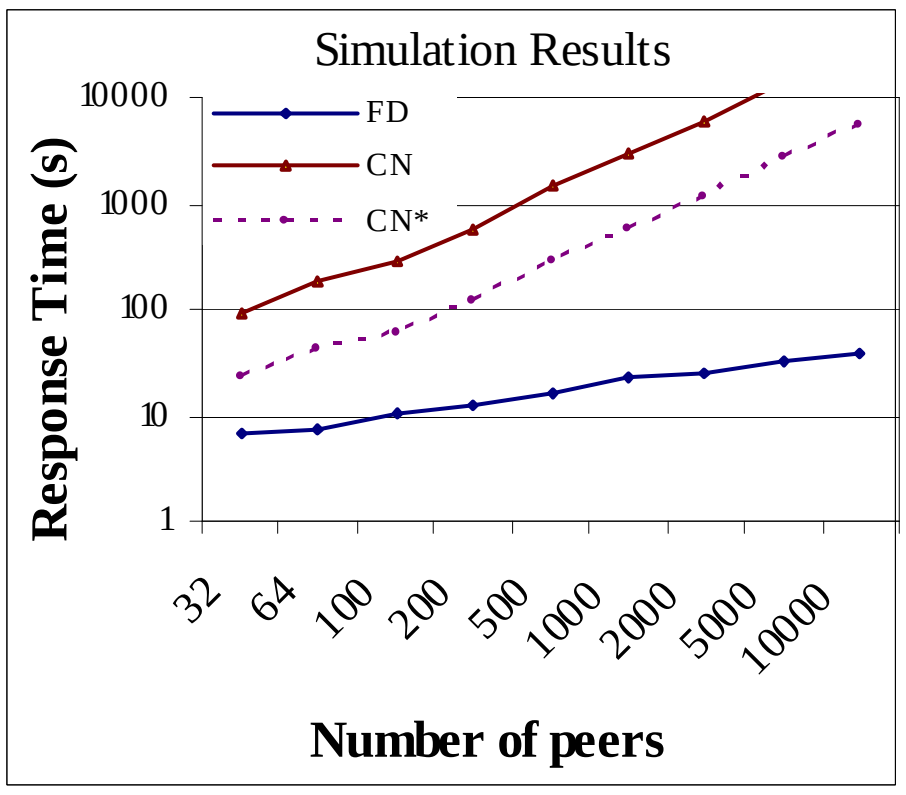

**Fig. 3.** Response time vs. number of peers

### Effect of Latency and Bandwidth

In this section, we study the effect of latency and bandwidth on response time. We only use our simulator since these parameters are fixed in our implementation. In the previous simulation tests the latency and bandwidth were normally distributed random numbers with mean values of 200 (ms) and 56 (kbps) respectively. In this test, we vary the mean values of the latency and bandwidth and study their effects on the response time.

Figure 4 shows how response time decreases with increasing bandwidth, with the other simulation parameters set as in Table 1. Increasing the bandwidth has strong, similar effect on all three algorithms which have to transfer some data over the network. FD outperforms the other algorithms for all the tested bandwidths.

Figure 5 shows how response time evolves with increasing latency, with the other simulation parameters set as in Table 1. Latency has little effect on the CN algorithm, because the peers return their results directly to the query originator, and do not bubble up the results. Although FD outperforms the other algorithms for all the tested values, high latency, *e.g.* more than 1000 ms, has strong impact and increases response time much. However, below 1000 ms, latency has not much effect on FD's response time. According to studies reported in [18], more than 80% of links between peers have good latency, less than 280 ms, for which FD has very good performance.

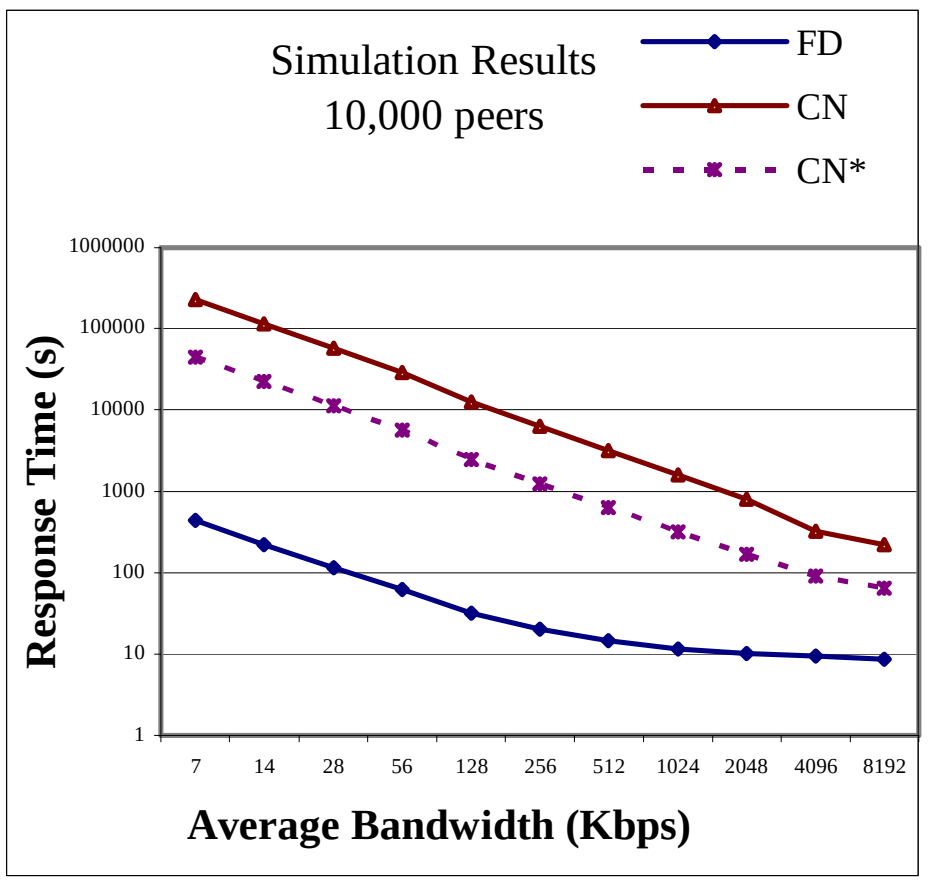

**Fig. 4.** Effect of bandwidth on response time

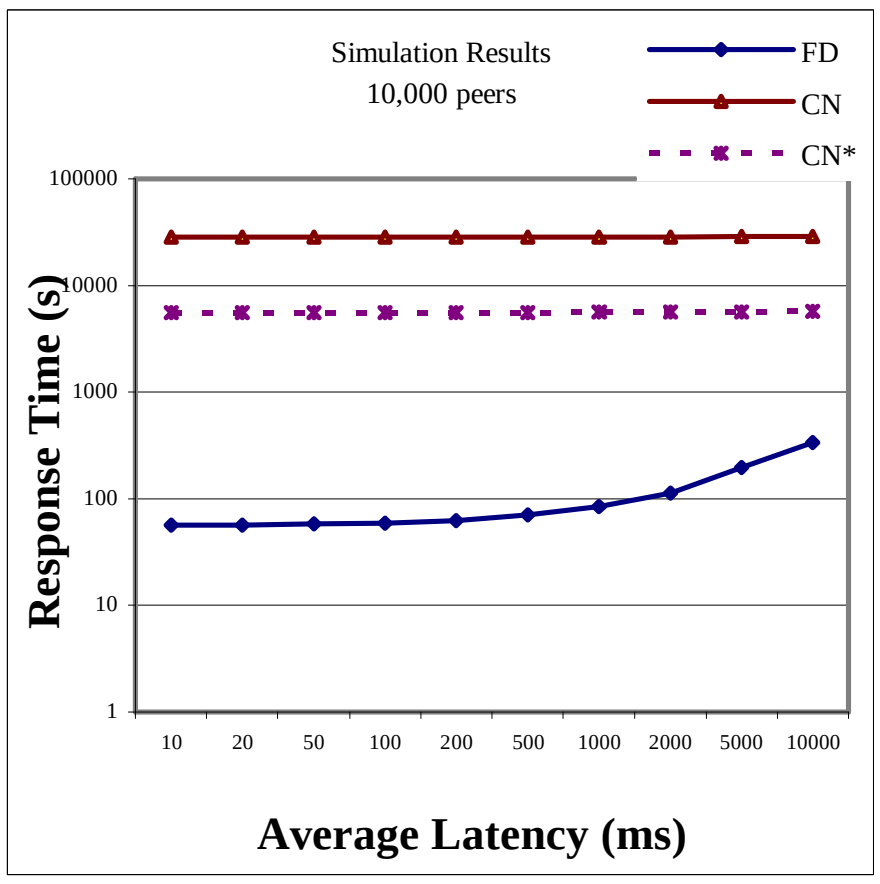

**Fig. 5.** Effect of latency on response time

## 5.3 Communication Cost

Now, using our simulator we study the communication cost of FD in its basic form and also with the strategies proposed in Section 3.2 for reducing the communication cost. We measure the communication cost in terms of the number of bytes, which should be transferred on the network for processing a top-n query *Q*.

**Effect of Strategy 1 and Strategy 2**

In this section, we study the communication cost of three versions of FD: 1) its basic form, noted as FD-Basic; 2) using Strategy 1, denoted by FD-St1; 3) using a combination of Strategy 1 and Strategy 2, denoted by FD-Str1+2.

Figure 6 shows how communication cost evolves with increasing the number of peers, with the other parameters set as in Table 1. With 10,000 peers, the communication cost of FD-Basic is about 5 MB, but FD-Str1+2 has reduced this cost to about 3.5MB, thus approximately 30% reduction in communication cost.

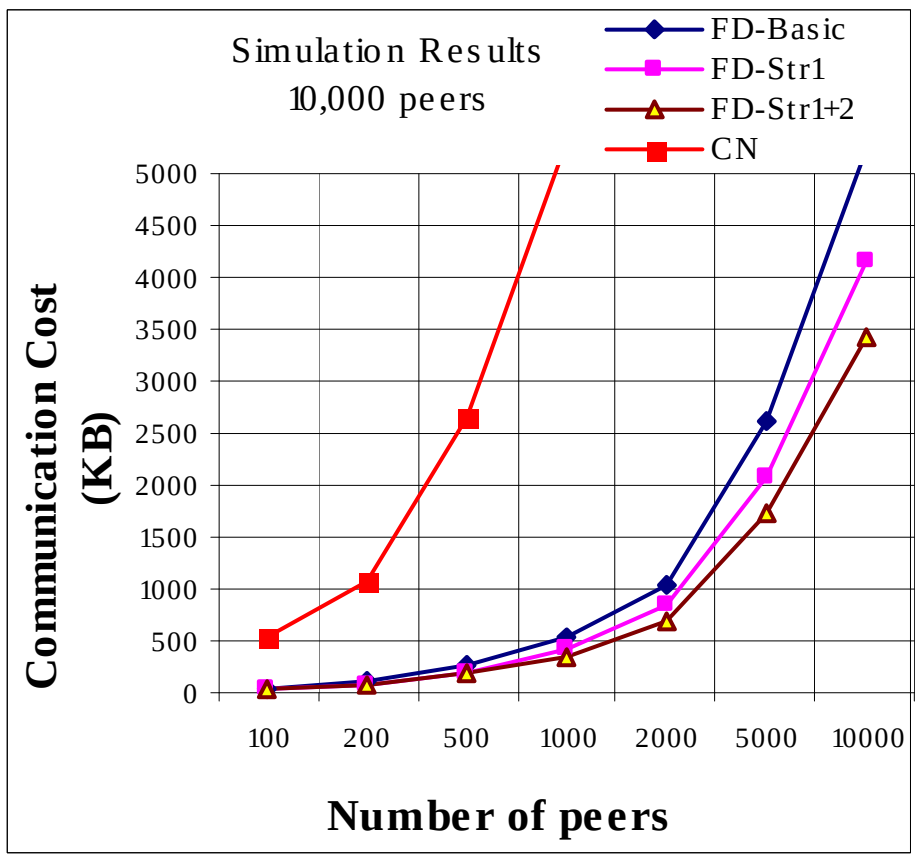

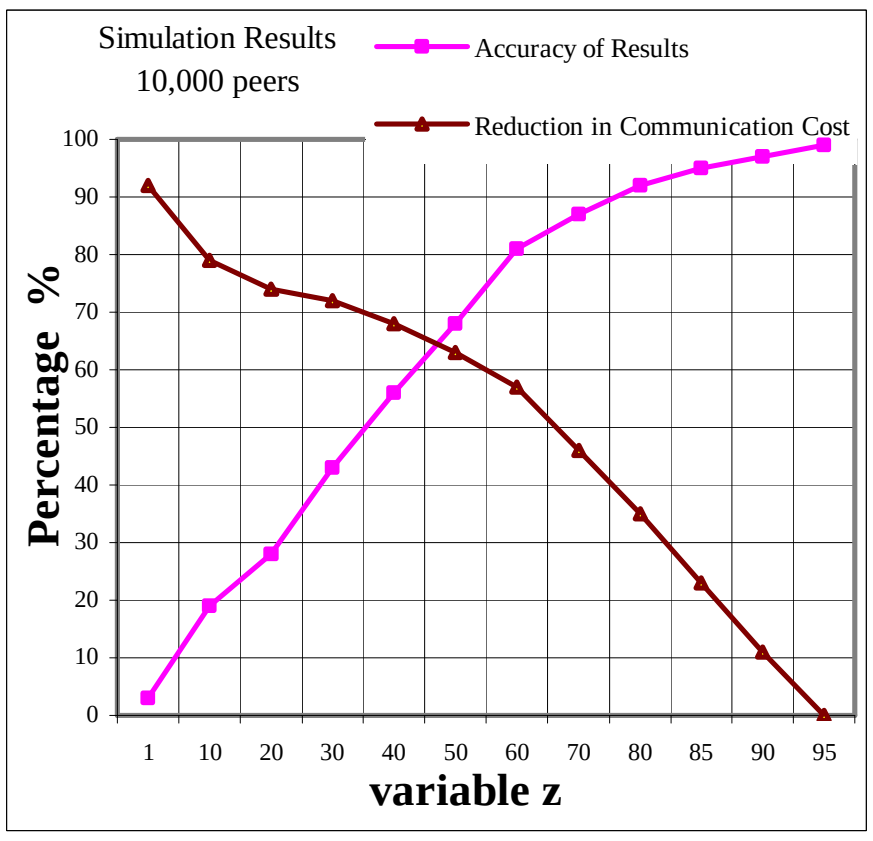

**Fig. 6.** Communication cost vs. number of peers

**Fig. 7.** Effect of using statistics for selecting "best" neighbors

**Effect of using statistics on communication cost**

As stated in Section 3, we can use the statistics gathered during previous query executions in order to select a subset of the "best" neighbors, which are more likely to return top results, and to send $Q$ only to them, thereby reducing the communication cost. In this section, we study the effect of this strategy on the communication cost. We measure the communication cost in terms of the number of bytes which should be transferred over the network for processing a Top-k query $Q$.

We used the following heuristic for selecting the neighbors: send $Q$ to the neighbors which the position of their greatest score in the merged score-list was lower than $z \times n$, where $z \leq 1$.

We also studied the effect of the above heuristic on the *accuracy* of the returned results. We define the accuracy of results as follows. Let $P_Q \subseteq P$ be a set containing the query originator and all peers that receive $Q$. Let $T_Q$ be the set of the $k$ top results owned by the peers involved in $P_Q$. Let $T_r$ be the set of the results which will be returned to the user as the response of $Q$. We denote the accuracy of results by $ac_Q$ and we define it as $ac_Q = (|T_Q \cap T_r|) / |T_Q|$.

We increased $z$ from zero to one and we measured $ac_Q$ and the percentage of the reduction in communication cost. The results are depicted in Figure 7. For $z = 0.80$, the accuracy of results is higher than *90%*, despite the fact that the communication cost is reduced by approximately *35%*. Thus, with a small loss of accuracy, we can obtain a significant reduction in communication cost.

## 5.4 Dealing with Peers' Dynamicity

Using the simulator we now study the effect of the strategies presented in Section 4 for dealing with the dynamic behavior of P2P systems, on the accuracy of results.

We defined the accuracy of results $ac_Q$ in Section 5.3. We now measure $ac_Q$ in two versions of FD. The first version, denoted as *FD-Basic*, is the basic form of FD in which peers discard the late score-lists, as well as their merged score-list in the case that their parent is inaccessible. The second one, denoted as *FD-Dynamic*, is a version in which in the case of receiving late score-lists or inaccessibility of the parent, the peers use the urgent score-lists for bubbling up the late score-list or merged score-list as described in Section 4.

In our tests, we investigated how the accuracy of results changes with varying the average lifetime of the peers according to the distributions observed in [18]. The lifetime of a peer is defined as the time period the peer stays in the P2P system. In our tests, we assumed that the query originator doesn't leave the system before releasing the results to the user.

The results are shown in Figure 8. For lifetimes above 4 minutes, the accuracy of results in FD-Dynamic is approximately one. But, in FD-Basic, even for lifetimes above one hour, the accuracy of results is less than one. This shows the excellent impact of our strategy on the accuracy of the results in dynamic environments.

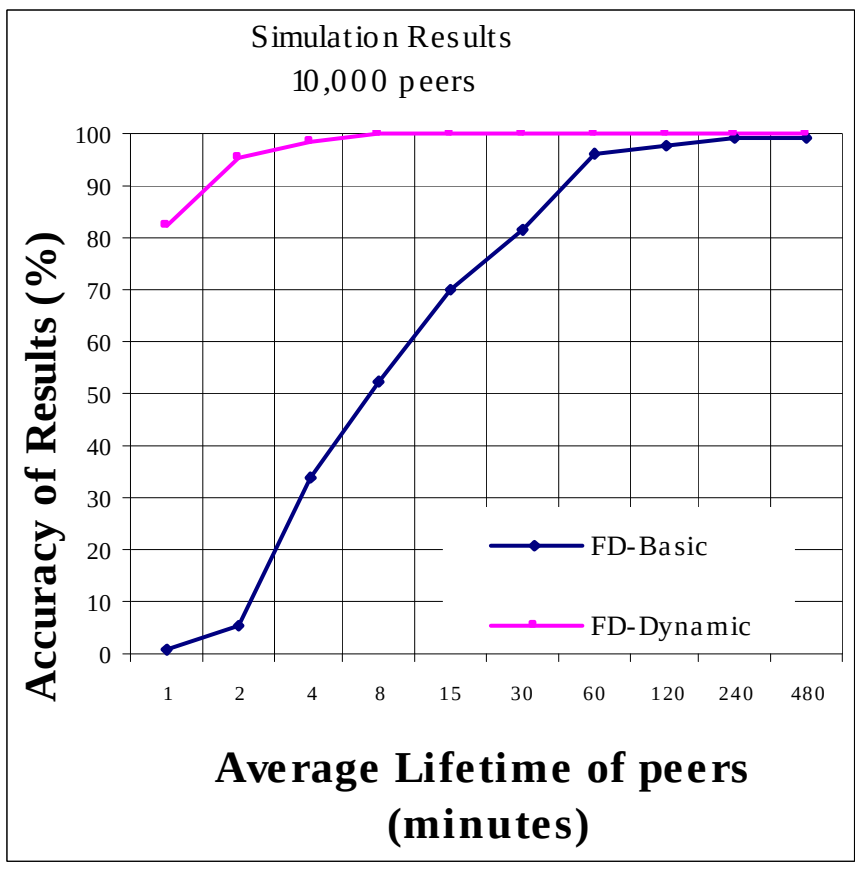

**Fig. 8.** Accuracy of results considering the dynamic behavior of P2P systems

## 6 Related Work

Most of the techniques proposed for Top-k query processing in distributed systems are based on histograms, maintained at a central site, to estimate the score of databases with respect to a query and send the query to the databases that are more likely to involve top results [24][25]. In [24], a two-step method for Top-k query processing in distributed systems with possibly uncooperative local systems is proposed. The first step determines which databases are likely to contain the top results and rank the databases with respect to the given query. The second step determines how the ranked databases should be searched and which tuples from the searched databases should be returned. A central node maintains some histograms to estimate the rank of the data-bases with respect to the query. However, this technique can not be used efficiently in a P2P environment, because keeping histograms up-to-date with autonomous peers that may join or leave the system at any time is difficult.

In the context of P2P systems, little research has concentrated on the Top-k query processing. In [21], the authors present a Top-k query processing algorithm for Edutella, a super-peer network in which a small percentage of nodes are super-peers and are assumed to be highly available with very good computing capacity. The super-peers are responsible for Top-k query processing and other peers only execute the queries locally and score their resources. Although very good for super-peer systems, this technique cannot apply efficiently to other P2P systems, in particular, unstructured, since there may be no peer with higher reliability and computing power. In contrast, FD makes no assumption about the P2P network topology and the existence of certain peers.

PlanetP [8] is a P2P system that constructs a content addressable publish/subscribe service using gossiping to replicate global documents across P2P communities up to ten thousand peers. In PlanetP, a Top-k query processing method is proposed that works as follows. Given a query $Q$, the query originator computes a relevance ranking of peers with respect to $Q$, and then contacts them one by one from top to bottom of ranking and asks them to return a set of their top-scored document names together with their scores. To compute the relevance of peers, a global fully replicated index is used that contains term-to-peer mappings. In a large P2P system, keeping up-to-date the replicated index is a major problem that hurts scalability. In contrast, our algorithm does not use any replicated data.

For the cases where a data item can have multiple scores at different sites, *e.g.* the amount of a customer's purchase in several stores, the Threshold Algorithm (TA) for monotonic score aggregation [9] stands out as an efficient method. There have been many works in order to optimize the TA algorithm in terms of communication cost and response time, *e.g.* [22] and [13]. In our case, each data item has a unique score. However, we could also use variations of the TA algorithm in the case of multiple scores.

## 7 Conclusion

In this paper, we proposed FD, a fully distributed framework for executing Top-k queries in unstructured P2P systems, with the objective of reducing network traffic. FD requires no global information, does not depend on the existence of certain peers, reduces significantly the communication cost, and addresses the volatility of peers during query execution.

We validated the performance of FD through implementation over a 64-node cluster and simulation using the BRITE topology generator and SimJava. The experimental and simulation results show that FD has logarithmic scale up. The simulations also show the excellent performance of FD, in terms of communication cost and response time, compared with two baseline algorithms. The results show that our strategies can reduce the communication cost significantly, *i.e.* by about 35%. They also show that, by selecting a good heuristic and well adjusting the parameters, we can take advan-tage of statistics to achieve a significant reduction, *i.e.* of more than 35%, in communication cost without a significant loss in accuracy, less than 10%. Furthermore, the simulation results show that the algorithms, proposed in FD for addressing the dynamic behavior of P2P systems, are effective.

As future work, we plan to deal with replicated data in P2P Top-k query processing. In the case of data replication, with our algorithm, there may be replicated data items in the final score-list. This may be fine for the user as it is an indication of the items' usefulness (in a P2P system, the most useful data get most replicated). But we could also identify replicated items. A simple solution is to add information in the score-lists to help eliminate duplicates (*e.g.* key values for relational data, descriptors for documents). An issue then is to optimally choose the replicas to access.

## Appendix A: Computing the Wait time

In the Local Query Execution phase (see Section 3), each peer *p*, after sending *Q* to its neighbors and executing *Q* locally, must wait to receive the results of its neighbors. However, some of the neighbors may leave the P2P system and never send any result to *p*. Thus, we must determine a limit for *p*'s wait time.

Let *ttl* be the value of *TTL* when *p* sends *Q* to its neighbors. We define *Response(Q,ttl)* as the time for *p* to receive all its neighbors' results. Thus, *Response(Q,ttl)* is the optimal value for *p's* wait time. If *p* waits less than *Response(Q,ttl)*, it will lose the results of some neighbors. If it waits more time, it will increase the overall response time. Therefore, we estimate *Response(Q,ttl)* in order to use it for setting the wait time of *p*. *Response(Q,ttl)* is made of the following cost components.

1. The *forward time*, which is the time to send *Q* from *p* to its *descendants, i.e. p*'s neighbors, *p*'s neighbors' neighbors, and so on, until TTL reaches zero.
2. The *local query execution time*, which is the time needed for the local execution of *Q* by the descendants of *p*.
3. The *merge time*, which is the time it takes the descendants of *p* merge their local scores with the received score-lists.
4. The *backward time*, which is the time it takes the descendants of *p* to bubble up their score-lists, until they reach to *p*.

To estimate these four cost components, we consider the maximum number of tasks, which must be done sequentially. Because the value of *TTL* of *Q* when *p* sends it to the neighbors is *ttl*, the forward time has at most *ttl* sequential sendings of *Q*. The local execution of *Q* can be done in parallel by all its neighbors, thus we consider only one sequential local query execution. To estimate the merge time, we consider the maximum number of peers that must do their merge operation sequentially which is *ttl-1* (only the peers that receive *Q* with *TTL=1* have not to do a merge operation).

The backward time consists of at most *ttl* sequential sendings of score-lists. Using the cost parameters described in Table 2 which we discuss below, we can state that:

$$Response(Q,ttl) \leq ttl \times T_{Qsnd}(Q) + T_{exec}(Q) + ttl \times T_{SLsnd}(k) + (ttl - 1) \times T_{Merge}(k) \quad (1)$$

Now, let $Wait_p(Q,ttl)$ be the time that peer *p* must wait after sending the query *Q* with $TTL=ttl$ to its neighbors, we can set $Wait_p(Q,ttl)$ as follows:

$$Wait_p(Q,ttl) = ttl \times T_{Qsnd}(Q) + T_{exec}(Q) + ttl \times T_{SLsnd}(k) + (ttl - 1) \times T_{Merge}(k) \quad (2)$$

**Table 2.** Cost parameters for estimating the wait time

| Parameter | Description |
|---|---|
| $T_{Qsnd}(Q)$ | Maximum time needed to send *Q* from a peer to its neighbor. |
| $T_{exec}(Q)$ | Maximum time to execute *Q* locally. |
| $T_{SLsnd}(k)$ | Maximum time to send a score-list containing *k* scores (and addresses) from a peer to its parent. |
| $T_{Merge}(k)$ | Maximum time for a peer to merge all score-lists received from its neighbors with its local top scores. Each score-list contains at most *k* couples. |

Formula (2) relies on the cost parameters described in Table 2. To discuss how we can obtain these cost parameters, we classify them into network-dependent parameters and local processing parameters.

**Network-dependent parameters**, *i.e.* $T_{Qsnd}(Q)$ and $T_{SLsnd}(k)$. These parameters depend on the P2P physical network characteristics, *e.g.* the latency and data transfer rate between peers. Peers can estimate these parameters using statistics gathered from previous query executions. During the bubbling up of the score-lists, we can also record some network characteristics like minimum latency and minimum data transfer rate between peers. Using these statistics, *p* can simply compute the network dependent parameters.

**Local processing parameters**, *i.e.* $T_{exec}(Q)$ and $T_{Merge}(k)$. Estimating $T_{exec}(Q)$ precisely is hard because it depends on many parameters, *e.g.* the computing power of peers, their load, their database size, etc. Thus, we simply let the user give a threshold *T* for the maximum local query execution time. If a peer cannot execute *Q* within *T* time units, its result may be discarded. In fact, this threshold is the budget that the user invests for the local query execution. She can adjust this parameter according to the desired trade-off between result completeness and response time. For instance, a low value of *T* may result in loosing the top results of the peers that cannot execute *Q* within *T* time units. Estimating $T_{Merge}(k)$ is easier because it is a function of *k* and the number of *p*'s neighbors. Furthermore, the time to merge score-lists is typically much smaller than the other cost parameters. So a simple constant value can be easily computed.